\documentclass{aa}
\usepackage{graphicx}
\usepackage{longtable}
\usepackage{txfonts} 

\newcommand{\dg}{\ensuremath{^\circ}}
\newcommand{\hess}{H.E.S.S.$\,$}

\begin{document}

\title{Very high energy gamma rays from the composite SNR G\,0.9+0.1}

\author{F. Aharonian\inst{1}
 \and A.G.~Akhperjanian \inst{2}
 \and K.-M.~Aye \inst{3}
 \and A.R.~Bazer-Bachi \inst{4}
 \and M.~Beilicke \inst{5}
 \and W.~Benbow \inst{1}
 \and D.~Berge \inst{1}
 \and P.~Berghaus \inst{6} \thanks{Universit\'e Libre de 
 Bruxelles, Facult\'e des Sciences, Campus de la Plaine, CP230, Boulevard
 du Triomphe, 1050 Bruxelles, Belgium}
 \and K.~Bernl\"ohr \inst{1,7}
 \and C.~Boisson \inst{8}
% ...
}

\offprints{J.A. Hinton, \email{Jim.Hinton@mpi-hd.mpg.de}}

\institute{
Max-Planck-Institut f\"ur Kernphysik, P.O. Box 103980, D 69029
Heidelberg, Germany
\and
 Yerevan Physics Institute, 2 Alikhanian Brothers St., 375036 Yerevan,
Armenia
\and
University of Durham, Department of Physics, South Road, Durham DH1 3LE,
U.K.
\and
Centre d'Etude Spatiale des Rayonnements, CNRS/UPS, 9 av. du Colonel Roche, BP
4346, F-31029 Toulouse Cedex 4, France
\and
Universit\"at Hamburg, Institut f\"ur Experimentalphysik, Luruper Chaussee
149, D 22761 Hamburg, Germany
\and
Physique Corpusculaire et Cosmologie, IN2P3/CNRS, Coll{\`e}ge de France, 11 Place
Marcelin Berthelot, F-75231 Paris Cedex 05, France
\and
Institut f\"ur Physik, Humboldt-Universit\"at zu Berlin, Newtonstr. 15,
D 12489 Berlin, Germany
\and
LUTH, UMR 8102 du CNRS, Observatoire de Paris, Section de Meudon, F-92195 Meudon Cedex,
France
% ...
}

\date{Received ? / Accepted ?}

\abstract{

Very high energy ($>$100~GeV) gamma-ray emission has been detected 
for the first time from the 
composite supernova remnant G\,0.9+0.1 using the \hess\ instrument.
The source is detected with a significance of $\approx$$13\,\sigma$, 
and a photon flux above 200~GeV of 
($5.7\pm0.7_{stat}\pm1.2_{sys})\,\times\,10^{-12}$
cm$^{-2}$s$^{-1}$,
making it one of the weakest sources ever detected at TeV energies.
The photon spectrum is compatible with a power law ($dN/dE \propto E^{-\Gamma}$) with
photon index $\Gamma = 2.40\pm0.11_{stat}\pm0.20_{sys}$.
The gamma-ray emission appears to originate in the plerionic core of
the remnant, rather than the shell, and can be plausibly explained as 
inverse Compton scattering of relativistic electrons.

;

}

\maketitle

%_________________________________________________________________
\section{Introduction}
\label{intro}

G\,0.9+0.1 is a well known composite supernova remnant, recognised as
such from its radio morphology (Helfand~\&~Becker~\cite{Helfand}).
It exhibits a bright compact core ($\sim$$2'$ across) surrounded by an
8$'$ diameter shell. The radio spectrum of the core is significantly
harder ($\alpha$ $\approx$ 0.12) than that of the shell ($\alpha$
$\approx$$-0.77$) (LaRosa et~al.~\cite{LaRosa}).  X-ray observations
of the nebula with BeppoSAX (Mereghetti et~al.~\cite{BeppoSAX}),
Chandra~(Gaensler et~al.~\cite{Chandra}) and XMM-Newton~(Porquet
et~al.~\cite{XMM}) have unambiguously identified the core region as a
pulsar wind nebula (PWN).  A plausible candidate for the central pulsar
is the hard spectrum point source CXOU J174722.8$-$280915; however
no pulsed emission has been detected. Observations
with XMM have revealed a softening of the X-ray spectrum with
increasing distance from the centre of the PWN, a signature of energy
loss of electrons within the nebula. The location of G\,0.9+0.1 in the
Galactic Centre (GC) region suggests a distance of $\sim$8.5~kpc
(Mezger~et~al.~\cite{GC}). At this distance the size of the shell
implies an age of a few thousand years for the SNR.
At gamma-ray energies G\,0.9+0.1 has not been detected previously.  
It is not associated with any EGRET source and
the only published observation in the TeV domain is an unconstraining
upper limit from the HEGRA collaboration (flux $<$$4.6\times\,10^{-12}$
cm$^{-2}$s$^{-1}$ above 4.2~TeV, Aharonian et~al.~\cite{HEGRA}).

The High Energy Stereoscopic System (H.E.S.S.) is a newly 
completed instrument designed to study astrophysical 
gamma radiation in the energy range 100 GeV -- 10 TeV.
H.E.S.S. consists of four imaging atmospheric Cherenkov 
telescopes (Hinton~\cite{HESS};
Bernl\"ohr et~al.~\cite{HESSOptics}; Vincent et~al.~\cite{HESSCamera};
Aharonian et~al.~\cite{HESSCalib}); shower images are required
in at least two of them to trigger the detector~(Funk et~al.~\cite{HESSTrigger}).  
Using the stereoscopic technique the instrument reaches an angular
resolution of $\sim$$0.1^{\circ}$ and a point source sensitivity 
of $1$\% of the flux from the Crab Nebula ($5\,\sigma$ in 25 hours).

Observations of the GC region were made with the partially
complete \hess\ system in 2003. These data revealed emission 
from the Sgr\,A region, consistent with the position of
Sgr\,A$^{*}$ (Aharonian et~al.~\cite{HESSGalCen}).  Within the same
field of view indications for a signal from the supernova remnant 
G\,0.9+0.1 were seen at the $4\,\sigma$ level.
A relatively deep H.E.S.S. observation of this region with the
complete telescope system was performed in March--September 2004.  The
results from this dataset relevant to G\,0.9+0.1 are presented here. A
detailed discussion of the implications of the 2004 dataset for the
region around Sgr\,A$^{*}$ will be presented elsewhere.

\section{\hess\ Observations and Results}

The large field of view of the \hess\ cameras ($\sim$5$^{\circ}$)
provides good sensitivity for point sources at an
angular distance up to $\sim$$2^{\circ}$ from the pointing direction
of the telescope system.  The bulk of observations presented here were
taken in \emph{wobble mode} around the position of Sgr\,A$^{*}$. In
this mode 28-minute runs are taken 
pointing 0.5$^{\circ}$ away from the nominal source in
alternating directions. Approximately 20\% of the data were taken in
wobble mode around G\,0.9+0.1 itself. The observations have an
average offset of 0.9$^{\circ}$ from the SNR. 
The off-axis sensitivity of the system derived from
Monte Carlo simulations has been confirmed via observations of the
Crab Nebula (Aharonian et~al.~\cite{HESSCrab}).
The total live time of the observations after run selection 
is 50 hours, from a total observation time of 60 hours. 
The mean zenith angle of observation was
18$^{\circ}$, resulting in an energy threshold of 170~GeV
after standard cuts.

\begin{figure}[h]
\centering
\includegraphics[width=8.5cm]{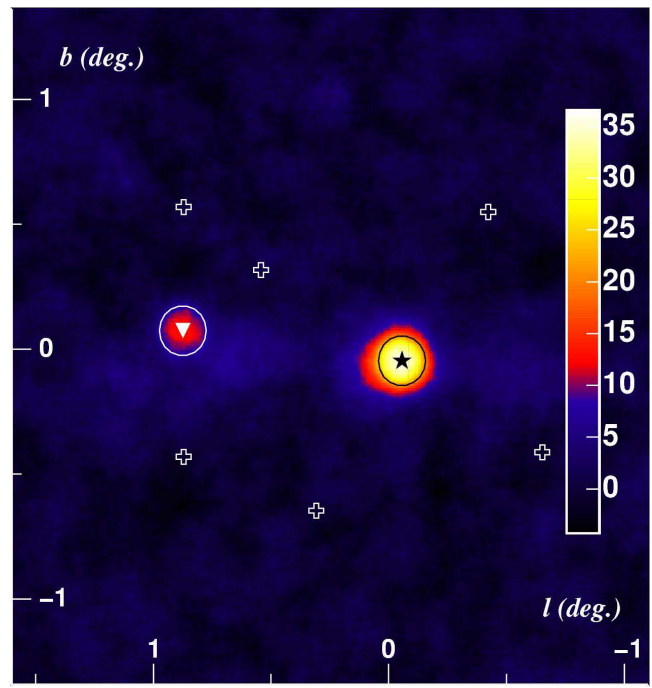}
\caption{
  Gamma-ray point source significance map for the region around the
  GC. The position of G\,0.9+0.1 is marked with a 
  triangle. Sgr\,A$^{*}$ is marked with a star. The circles
  show the integration regions contributing to the quoted 
  significance at the positions of these two objects. The six
  telescope pointings are shown as crosses.
} 
\label{fig:skymap}
\end{figure}

Standard event reconstruction was applied to these data: tail-cuts
image cleaning, Hillas parameterisation and stereoscopic direction
reconstruction based on the intersection of image axes~(Aharonian et
al.~\cite{HESS2155}). Cuts on the scaled width and length of images
(optimised on gamma-ray simulations and off-source data) are made 
to suppress the hadronic background. An additional cut on the minimum
size of images (200 photoelectrons) is made to further suppress the 
background and select a subset of events with a superior angular resolution
(0.07\dg rms), but with an increased energy threshold of 350~GeV.
Events with directions reconstructed
within an angle $\theta$ of a trial source location are considered
on-source where $\theta\,<\,0.1$\dg.  A ring of radius 0.5$^{\circ}$
and an area 7 times that of the on-source region
is used to derive a background estimate. The on-source and background
region counts, together with a normalisation factor
for the different acceptance of these regions, are used to
derive the statistical significance of any excess following the
likelihood method of Li~\&~Ma~(\cite{LiMa}).

Figure~\ref{fig:skymap} shows a significance sky map for the field of
view of the H.E.S.S. GC observations derived as described
above.  Two sources of very high energy (VHE) gamma-rays are clearly visible:
the Galactic Centre (HESS\,J1745$-$290) and a second source (HESS\,J1747$-$281) 
coincident with the SNR G\,0.9+0.1.
The statistical significance of HESS\,J1745$-$290 is
$35\,\sigma$ in this dataset, compared with
$11\,\sigma$ in our 2003 data (Aharonian et~al.~\cite{HESSGalCen}). 
This increase is expected (for a steady
source) from the greater sensitivity of the full 4-telescope array
with respect to the 2-telescope configuration used in 2003, and from
the increased exposure time in 2004.

The H.E.S.S. 2003 dataset showed evidence for gamma-ray emission 
at the position of G\,0.9+0.1 at the $4\,\sigma$ level. 
This evidence is confirmed by the presence of a $13\,\sigma$ excess 
at the same position in the independent 2004 dataset.
The gamma-ray like excess is well 
fit by the point spread function of the instrument, yielding a best
fit position of $l$ = 0.872$^{\circ}\,\pm\,0.005^{\circ}$, 
$b$ = 0.076$^{\circ}\,\pm\,0.005^{\circ}$, consistent within
statistical errors with the position of the PWN in G\,0.9+0.1
($l$ = 0.871$^{\circ}$, $b$ = 0.0772$^{\circ}$)~(Gaensler et~al.~\cite{Chandra}).
The systematic pointing error is approximately $20''$ in 
each direction for these data.

\begin{figure}
\centering
\includegraphics[width=8.5cm]{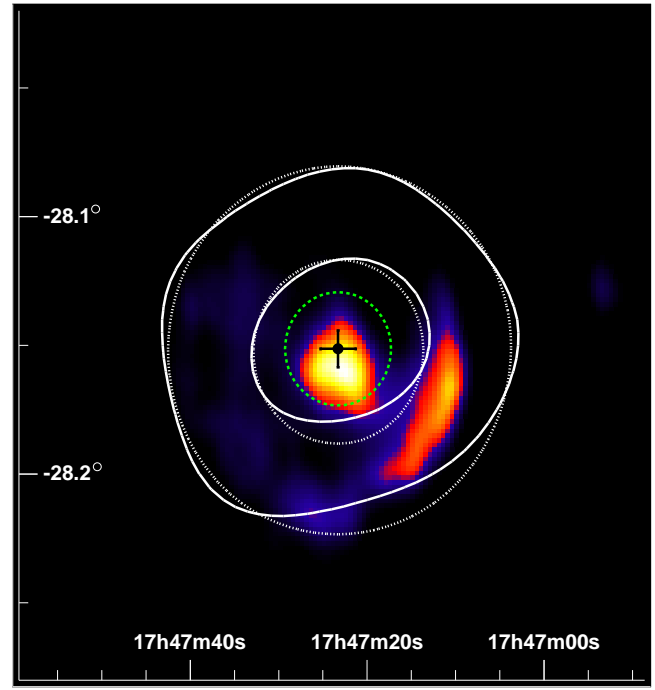}
\caption{
  90~cm radio flux map of G\,0.9+0.1 from LaRosa
  et~al.~(\cite{LaRosa}) (colour scale), overlayed
  with contours (at 40\% and 80\% peak brightness) of the 
  smoothed and acceptance corrected count map of gamma-ray candidates
  (solid lines).  The H.E.S.S. data are smoothed with a Gaussian of rms
  0.03$^{\circ}$ to reduce statistical fluctuations.
  The simulated point-spread function of the instrument is shown
  with dotted lines (also 40\% and 80\%). The best fit position of the VHE gamma-ray
  excess is shown with error bars showing combined statistical and
  systematic errors. The innermost circle (dashed) illustrates 
  the 95\% confidence limit on the rms size of the emission region.
} 
\label{fig:zoom}
\end{figure}

The simulated point-spread function of the instrument
provides an adequate description of the excess. 
Assuming a radially symmetric Gaussian emission region 
($\rho \propto \exp(-\theta^{2}/2\sigma_{\mathrm{source}}^2)$)
a 95\% confidence limit on the extension 
$\sigma_{\mathrm{source}}\,<\,1.3'$ is obtained.
For emission from a uniform thin shell a limit of $2.2'$ on the shell 
radius can be derived at the same confidence level. 
Figure~\ref{fig:zoom} shows a radio map of G\,0.9+0.1
at 90~cm (LaRosa et~al.~\cite{LaRosa}) showing the 
SNR shell and the central compact source (colour scale). 
Contours of VHE gamma-ray emission are superimposed along with the 
point spread function of the instrument for comparison.
The reconstructed gamma-ray source position is marked 
as a cross. Given the close positional coincidence, we identify the 
VHE gamma-ray source HESS\,J1747$-$281 with G\,0.9+0.1.

\begin{figure}
\centering
\includegraphics[width=8.2cm]{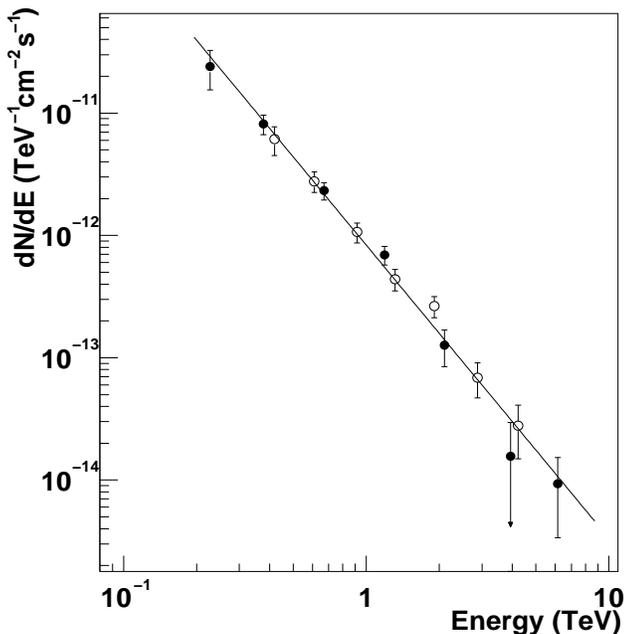}
\vspace{-0.2cm}
\caption{
  Reconstructed VHE gamma-ray spectrum of G\,0.9+0.1. 
  Empty circles show the spectrum derived using the rather
  tight cuts used for figures \ref{fig:skymap} and \ref{fig:zoom}. 
  Solid symbols show the spectrum derived using an image size cut of
  80 photoelectrons (optimised for maximal significance for 
  10\% Crab strength sources) rather than 200 photoelectrons.
  The line shows a power-law fit to the 80~pe data set.
}
\label{fig:spectrum}
\end{figure}

Figure~\ref{fig:spectrum} shows the reconstructed gamma-ray spectrum
of G\,0.9+0.1 using two sets of event selection cuts. A size cut of
80 photoelectrons (pe) is applied to extend the spectrum to lower energies.
This dataset can be fit by a power law in
energy with photon index $2.40\pm0.11_{stat}\pm0.20_{sys}$
and a flux above 200~GeV of 
($5.7\pm0.7_{stat}\pm1.2_{sys})\,\times\,10^{-12}$
cm$^{-2}$s$^{-1}$. 
The fit has a $\chi^2$/d.o.f. of 3.5/5. 
Using the harder image size cut of 200~pe used elsewhere in this
paper yields consistent results (photon index $2.29\pm0.14_{stat}$ and a flux of 
$(5.5\pm0.8_{stat})\,\times\,10^{-12}$ cm$^{-2}$s$^{-1}$ above 200~GeV).
This flux represents only 2\% of the flux from the Crab Nebula
above the 200~GeV spectral analysis threshold.
Although G\,0.9+0.1 is one of the weakest sources ever detected 
at TeV energies, the detection presented here is statistically
highly significant ($13\,\sigma$), 
demonstrating the sensitivity of the H.E.S.S. instrument. 

\section{Discussion}

X-ray emission from the shell of G\,0.9+0.1 is only marginally
detected above the local GC diffuse emission with a flux of
$2\times 10^{-12}$~erg~$\textrm{cm}^{-2}\,\textrm{s}^{-1}$ in the range
2--10~keV, and is consistent with both a thermal or non-thermal
origin~(Porquet~et~al.~\cite{XMM}).  The non-thermal flux from the PWN is
$5.8\times10^{-12}$~erg~$\textrm{cm}^{-2}\,\textrm{s}^{-1}$ in the
same energy range, corresponding to a luminosity of $\sim$$5\times
10^{34}$ erg\,s$^{-1}$ assuming a distance of 8.5~kpc.
The lack of strong non-thermal X-ray emission from the shell and the
point-like nature of the H.E.S.S. detection argue against an origin of
the VHE emission in the shell. Thus the PWN appears to be a more
compelling source for the VHE gamma-rays.

The total power radiated by G\,0.9+0.1 in VHE gamma-rays is $\sim$
$2$$\times 10^{34}$ erg\,s$^{-1}$ (at 8.5~kpc), compared with
$\sim$$4\times 10^{34}$ erg\,s$^{-1}$ for the Crab Nebula (at 2~kpc)
in the energy band $0.2$--$10$~TeV. Until now the Crab Nebula
represented the only firm
identification of VHE emission from a nebula powered by a young
neutron star.

A simple one-zone inverse Compton model for the VHE gamma-ray emission
(Kh\'elifi~\cite{Bruno}) can be used to explain the observed data. A
parent population of accelerated electrons with a broken power-law
spectrum is assumed. The magnetic field strength $B$ is assumed
to be uniform within the PWN and is a free parameter. 
The maximum energy of the electron spectrum is
kept fixed at $500$~TeV as it can not be derived due to the absence of
measurements in the hard X-ray/soft $\gamma$-ray band.
The synchrotron and inverse Compton (IC) emission of these 
electrons are calculated according to Blumenthal~\&~Gould~(\cite{Blumental}). 
The Klein-Nishina effect on the IC cross section is taken into
account. 

For sources in the GC region the presence of radiation densities
considerably higher than typical Galactic Plane values leads
to substantially enhanced IC emission (de~Jager~et~al.~\cite{DeJager}).
The seed photons for IC scattering are the cosmic microwave background
radiation (CMBR), the galactic diffuse emission around
$100\,\mu\textrm{m}$ from dust and around $1\,\mu\textrm{m}$ from
starlight. The energy spectra of the dust and starlight components
are taken from a recent estimation of the interstellar radiation field
(ISRF) used for the GALPROP code (Strong et~al.~\cite{GALPROP}). 
To account for possible local variations of the starlight component 
the energy density of this component is kept as a free parameter. 
The energy density of the dust ISRF is
kept fixed at $0.23\textrm{\,eV\,cm}^{-3}$ as used in GALPROP. 
The starlight ISRF energy
density, the electron spectrum parameters and $B$ are adjusted such
that the computed synchrotron and IC spectra match 
the observations.

\begin{figure}
\centering
\vspace{-0.3cm}
\includegraphics[width=9.5cm]{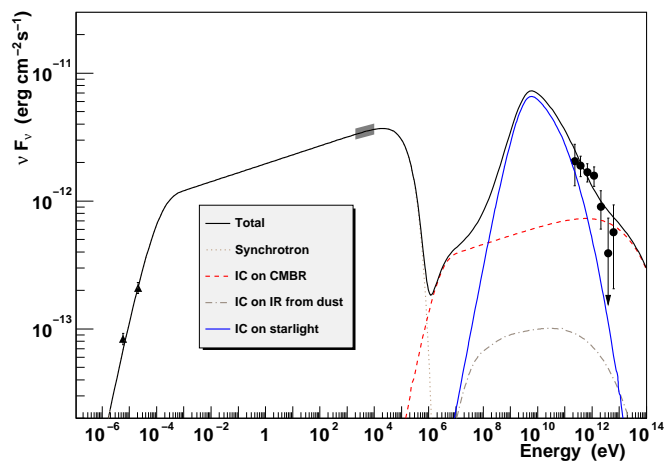}
\caption{The spectral energy distribution of the PWN in G\,0.9+0.1 from radio to
  VHE gamma-ray. Radio data (triangles) are taken from
  (Helfand~\&~Becker~\cite{Helfand}), 
  X-ray data (shaded box) from
  Porquet et~al.~\cite{XMM}. 
  The circles show the H.E.S.S. data from this work. 
  The solid curve shows a fit of a one-zone inverse
  Compton model to all data. Contributions of the 
  CMBR, IR from dust and starlight photon fields to the 
  IC emission are shown.
}
\label{fig:sed}
\vspace{-0.2cm}
\end{figure}

Figure~\ref{fig:sed} shows the broad-band spectral energy distribution
(SED) of G\,0.9+0.1, together with the best model fit. 
The fitted mean magnetic field strength is $\approx$$6\,\mu$G.
The model electron spectrum has a low energy index of $0.6$ 
and a high energy index of $2.9$ with a break at $25$~GeV.  
The best fit energy density of the starlight ISRF is 
about $5.7\textrm{\,eV\,cm}^{-3}$, that is 50\% smaller than
the value used in GALPROP. This derived value of $B$ 
is very close to the equipartition magnetic field ($\approx$$5\,\mu$G),
calculated from the whole fitted electron spectrum 
and given a source of radius 1$'$ at 8.5~kpc.
Given the observed variation of the X-ray spectrum
within the PWN, this one zone model is clearly somewhat
simplistic. However, such a model, describing the average
properties of the PWN, can describe the available data
with reasonable physical parameters.
Future measurements in the GeV regime with GLAST 
(sensitivity $2\times10^{-12}$~erg\,s$^{-1}$cm$^{-2}$ at 1~GeV) 
are needed to resolve the ambiguity between the magnetic
energy density and that of target radiation fields in this region.
Alternative explanations based on a hadronic origin
of the VHE emission (for example via the decay of 
neutral pions produced in $p$--$p$ collisions) are not excluded.

The detection of VHE gamma-ray emission from G\,0.9+0.1 provides the
first direct evidence for the acceleration of very energetic
particles in this object. The position and point-like nature 
of the gamma-ray emission, combined with the broad band SED of G\,0.9+0.1,
strongly suggest an origin of the VHE emission in the compact central 
source (previously identified as a pulsar wind nebula).

This detection represents the first step towards the study of
an emerging population of VHE gamma-ray emitting PWN. 
The detection of new PWN with accurate TeV spectra and with 
morphological measurements will provide 
information key to understanding particle acceleration in the vicinity
of pulsars.

%_________________________________________________________________
\section*{Acknowledgements}

The support of the Namibian authorities and of the University of Namibia
in facilitating the construction and operation of H.E.S.S. is gratefully
acknowledged, as is the support by the German Ministry for Education and
Research (BMBF), the Max Planck Society, the French Ministry for Research,
the CNRS-IN2P3 and the Astroparticle Interdisciplinary Programme of the
CNRS, the U.K. Particle Physics and Astronomy Research Council (PPARC),
the IPNP of the Charles University, the South African Department of
Science and Technology and National Research Foundation, and by the
University of Namibia. We appreciate the excellent work of the technical
support staff in Berlin, Durham, Hamburg, Heidelberg, Palaiseau, Paris,
Saclay, and in Namibia in the construction and operation of the
equipment. We would also like to thank J. Lazio (NRL) for provision of
90~cm radio data and A. W. Strong for providing a parameterisation of 
interstellar radiation fields.

%_________________________________________________________________

\end{document}